# T35: a small automatic telescope for long-term observing campaigns

**Susana Martín-Ruiz**[1], **Francisco J. Aceituno**[1], **Miguel Abril**[1], **Luis P. Costillo**[1], **Antonio García**[1], **José Luis de la Rosa**[1], **Isabel Bustamante**[1,2], **Juan Gutiérrez-Soto**[1,3], **Héctor Magán**[1,4], **José Luis Ramos**[1], **Marcos Ubierna**[1,5]

[1]*Instituto de Astrofísica de Andalucía - CSIC. Cno. Bajo de Huétor, 50. 18008 Granada, Spain*
[2]*Parque de las Ciencias. Avda de la Ciencia s/n. 18006 Granada, Spain*
[3]*GEPI, Observatoire de Paris, CNRS, Université Paris Diderot, place Jules Janssen, 92195 Meudon Cedex, France*
[4]*Centro Astronómico Hispano Alemán. C/ Jesús Durbán Remón 2-2, 04004 Almería, Spain*
[5]*SENER INGENIERIA Y SISTEMAS S.A. C/ Avda. Zugazarte, 56. 48930 Las Arenas – Bizkaia, Spain*

The T35 is a small telescope (14") equipped with a large format CCD camera installed in the Sierra Nevada Observatory (SNO) in Southern Spain. This telescope will be a useful tool for the detecting and studying pulsating stars, particularly, in open clusters. In this paper, we describe the automation process of the T35 and show also some images taken with the new instrumentation.

## 1. Introduction

At the beginning, the main motivation for carrying out the T35 project was the search for and study of pulsational behaviour of variable stars in open clusters. The role of open clusters, as stellar associations with a common origin, is fundamental in Astroseismology. The physical properties shared by the members of a cluster: distance, reddening, age and metallicity provide us very stringent constraints on the models, complementing the information obtained from the oscillation frequencies of the pulsating stars. Exhaustive studies on the incidence of variability and its behaviour, especially on pulsators located in the lower part of the Instability Strip (γ Doradus, δ Scuti or solar-type variables), help us to know better some of fundamental parameters (Teff, log$g$, chemical composition and rotational velocity) of these stars.

A previous systematic survey in search of γ Doradus variability in different open clusters with different metallicities and ages was performed between the years 1995 and 2000 [1-3]. More than 340 nights of observation at Sierra Nevada Observatory (SNO) (Granada, Spain), using photoelectric photometry in the Strömgren-Crawford system were used to carry out this study. Nine γ Doradus were found amongst the 41 variable stars detected in a sample of 175 members distributed among the 10 open clusters applying two methods based on different statistical tests to classify our light curves. The main outcomes were that the probability of finding γ Doradus stars increases if the sample is restricted to AF-type stars (effective temperature between 6900 and 7200 K), luminosity class IV-V (stars in the main sequence) and solar-type metallicity (Z=0.02) as well as that this probability was not bound to the age of the cluster but to its metallicity contradicting the theories published by other authors. Although our results were very fruitful due to the high precision of our uvbyβ measurements, i.e., less than two thousandth of magnitude, the number of member stars and clusters studied in the sample was small with the addition of entailing an enormous observational effort. Therefore, we needed a telescope of only modest aperture (30-40-cm) to reach the desired S/N in a reasonable time.

With this telescope it is possible to perform long-term observations of variable stars. Continuous observations in time as well as long time baseline campaigns are essential to the study of variable stars, including binary systems and pulsating stars. It is very difficult to be allocated long duration observing sessions on large telescopes.

The process of automation of the T35 has involved efforts in hardware, software and mechanics. A general block diagram of the final system is shown in Fig.3, which is described in detail in the next sections.

## 2. T35 setup

The T35 telescope (Fig. 1) is located at the Loma de Dílar (2896 m altitude), near the central building of Sierra Nevada Observatory (Granada, Spain). Before this 14" telescope was installed, the dome housed a different instrument, and therefore, first we had to restore and adapt the structure to our new instrumentation.

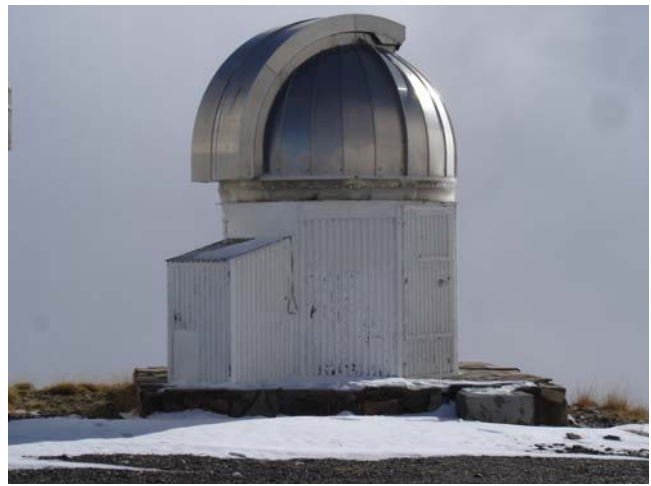

**Figure 1:** Dome of the T35 telescope

The 14" Schmidt-Cassegrain telescope (35.56-cm) is a Celestron CGE-1400. The telescope is equipped with an SBIG STL-11000 CCD Camera with a KAI-11000M CCD detector (4008 x 2762 pixels x 9 μm). Fig. 2 shows both instruments. The field of view is 31.70 x 21.14 arcmin with a scale of 0.2475 arcsec per pixel. The camera has a internal self-guiding camera Texas Instruments TC-237H (657 x 495 pixels x 7.4 μm) and an internal filter wheel with standard UBVRI Johnson - Cousins filters.

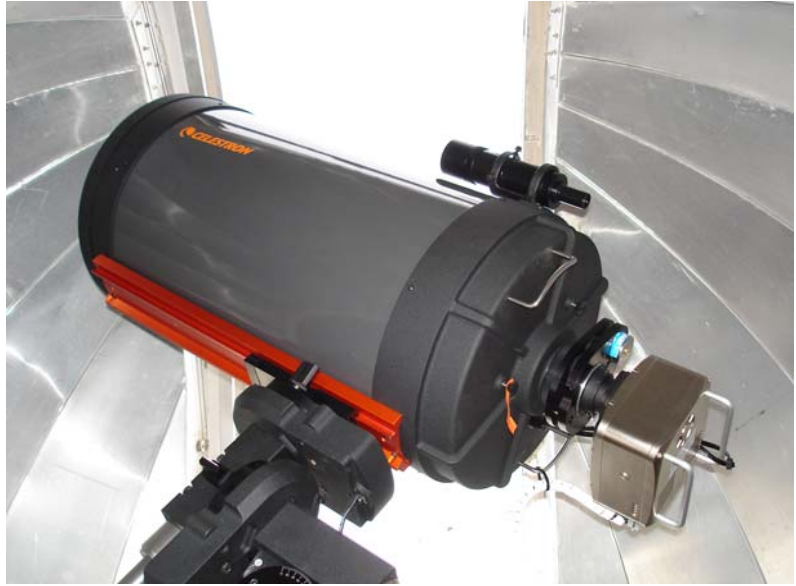

**Figure 2:** The telescope and CCD Camera

Since the beginning, the aim of our project was to install a telescope in order to perform long-term photometric observing campaigns. Owing to the OSN is a high mountain observatory where adverse weather conditions happen frequently, the T35 telescope had to work in remote mode with the greatest grade of autonomy. To date we are working to get this objective and hope to robotize the system in the near future.

Regardless of whether the mode of operation, fundamental requirements of pointing and tracking as well as a minimum precision in the photometric measurements are necessary. Table 1 show these basic parameters necessary for our objectives and those ones achieve in our telescope. The photometric accuracy, the 5-6 thousandth of magnitudes has been obtained observing a variable star during a non-photometric night, taking the frames with binning 3 x 3 and using faint comparison stars. Best outcomes, less than 2 mmag, can be achieved if the atmospheric conditions as well as the observing parameters (integration time, binning…) are optimum. Respect to the tracking accuracy, it can be improved using the internal self-guiding camera but the small size of chip does not often enable to find a bright star in the camera field of view. To solve this problem, it is being implemented an external system auto-guiding. More complicated is to obtain a higher value of pointing accuracy. The typical pointing values have been obtained using around 30 stars located in different positions in the sky. The telescope pointed out of 10' of arc in the 23% of the cases. To accurately point to objects, first we preformed the alignment procedure described in the telescope manual using two known stars. Despite the telescope was aligned properly, a pointing model was created to improve even more the pointing precision. This model has been made using the application *TPoint* which works quite well with the telescope control programme *TheSky Astronomy Software*. Although the observing target falls inside the field of view of the camera, its position can be correct by the user at the beginning of the observation.

|  | Scientific objectives | Achieved values |
|---|---|---|
| Pointing | Better than 5' of arc | 10' of arc |
| Tracking | 0" of arc in several min | 1."3 of arc in 2 min |
| Photometric accuracy | 1-2 thousandth of mag | 5-6 thousandth of mag |

**Table 1:** Values of pointing, tracking and photometric accuracy necessary for our scientific objectives and those achieved in the T35 telescope

The T35 telescope has been supported with funding from the Marie Curie Reintegration Grant `Detection and Survey of pulsating Star in Open Clusters: a step forwards in Asteroseismology' (MERG-CT-2004-513610) funded under the European Commission's Sixth Framework and to the Spanish project "Participación española en la misión espacial CoRoT" (ESP2004-03855-C03-01).

## 3. Control system

As mentioned above, the objective of this project was to automate and control remotely the movement of telescope and dome. The dome control system, in particular, is required to implement the following functions:
1. Continuous reading and updating of Azimut (Az).
2. Ability to move to an absolute position, from 0 to 359 degrees.
3. Ability to move to a relative position (+/- xxx degrees).
4. Ability to move to a prefixed park position.
5. Management of a zero cross reference position for loss of steps error detection.
6. Ability to modify remotely the constants of operation (park position, inertia constant, zero cross position…).
7. Measurements of consumption, and ability to detect a malfunction of the system through measurement of anomalous values.

As can be seen in Fig. 3, both telescope and dome can be controlled locally from a notebook or conventional PC installed inside the dome (Telescope Control System, hereafter TCS). However, the control of the system in usual operation is performed from one of the user computers in the main building of the observatory.

As regarding hardware elements, in this application both commercial (zero cross sensor, encoder, dome motor) and in-house developments (consumption measurement card, dome controller) have been used.

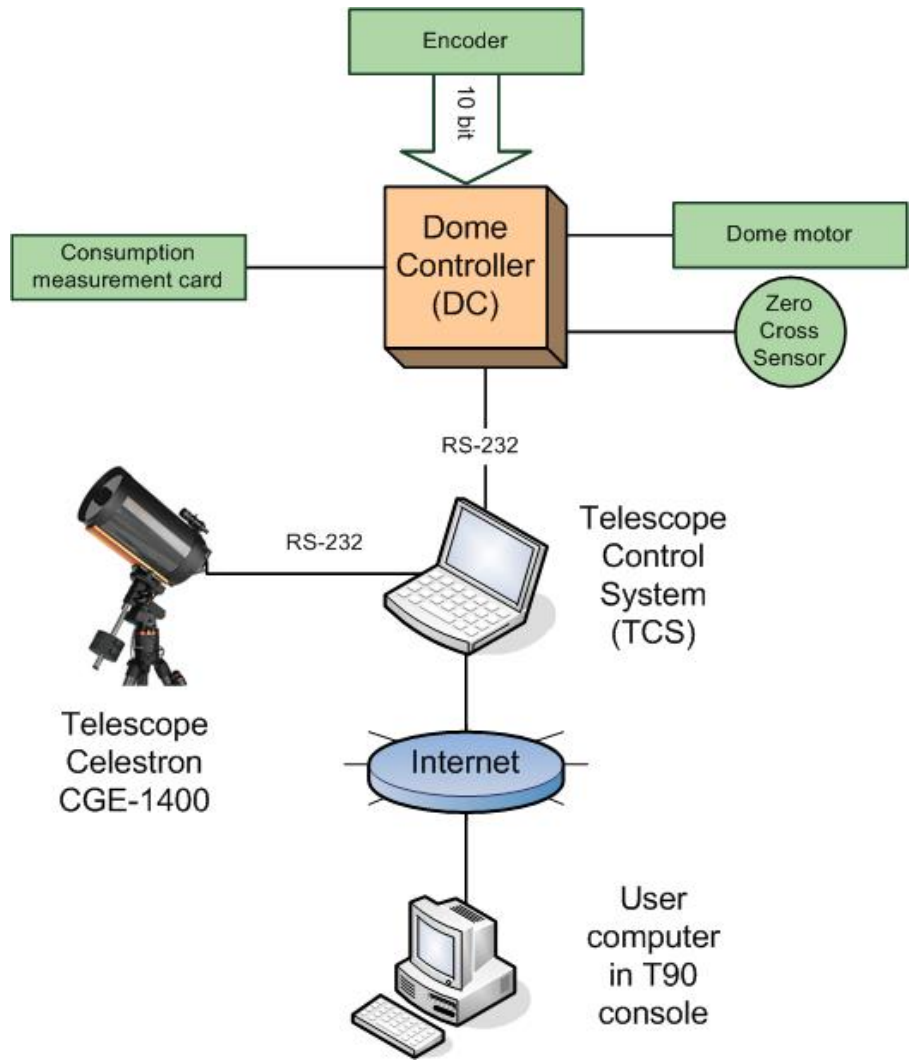

**Figure 3:** General block diagram of the system.

With respect to software, at least two programs were needed for this application, one to run on the Dome Controller (DC) and another on the TCS or user computer. Apart from these, an engineering program was also developed for technical purposes.

The main mechanical works were related to the design and fabrication of adaptors for the integration of the different elements in the dome, in order to assure a proper reading of the encoder and zero cross sensor minimizing loss of steps.

In the next sections, the main tasks, elements and programs developed for this application are discussed in more detail.

## 3.1. Hardware

The user interface and main control of both telescope and dome are provided by TCS, which is connected to Internet in order to allow the system to be controlled remotely. In practice, the system is controlled from the main user's computer on the T90 console. The TCS is connected via RS-232C to the telescope and DC.

The DC, developed specifically for this application, is based on a PIC18F458 microcontroller, in which we had experience from previous projects as regards programming and developing [4]. There exist several commercial modules for control of small and medium-sized domes. However, the solution chosen was based in a modular system designed in the *Instituto de Astrofísica de Andalucía* and used in previous projects. This solution presents several advantages over a commercial system: an in-house design is well known and documented, and therefore, it is easier to maintain. It allows for an exact adaption to particular requirements at a low price. The heart of the system, the microcontroller PIC18F458, can be easily programmed in high-level languages using different tools supplied by the manufacturer or third party providers. Finally, its modularity made it compatible and interchangeable with other modules used in our institution. These factors can make this DC attractive for other telescopes, provided that their requirements are similar to those of our system. In fact, the advantages associated to this in-house developed system (modularity, ease of programming, price) can result of interest for other institutions, not only to develop their own dome controller, but also other systems which requirements are affordable by the PIC18F458 microcontroller.

As can be seen in Fig. 3 to 5, the DC reads the dome position from an absolute Gray encoder with a resolution of 10 bits (Hohner, model CS10-81310311-1024). A zero cross sensor allows for loss of steps error detection. The system also employs a consumption reading card, used to detect and prevent breakdowns due to ice on the dome or malfunction of the motor.

In addition to the reading of encoder and zero cross sensor, the DC performs the actions on the dome motor. The DC uses a driver card for adaption of the control signals to the levels needed for acting on the dome motor. In order to facilitate the use of the previous operating hardware, the DC acts in parallel with buttons AZ+ and AZ-, which allow for manual movement of the dome. Consequently, there is no need for flipping between a manual and an automatic mode of operation. Fig. 3 to 6 show a general block diagram of the system and some pictures of its hardware implementation.

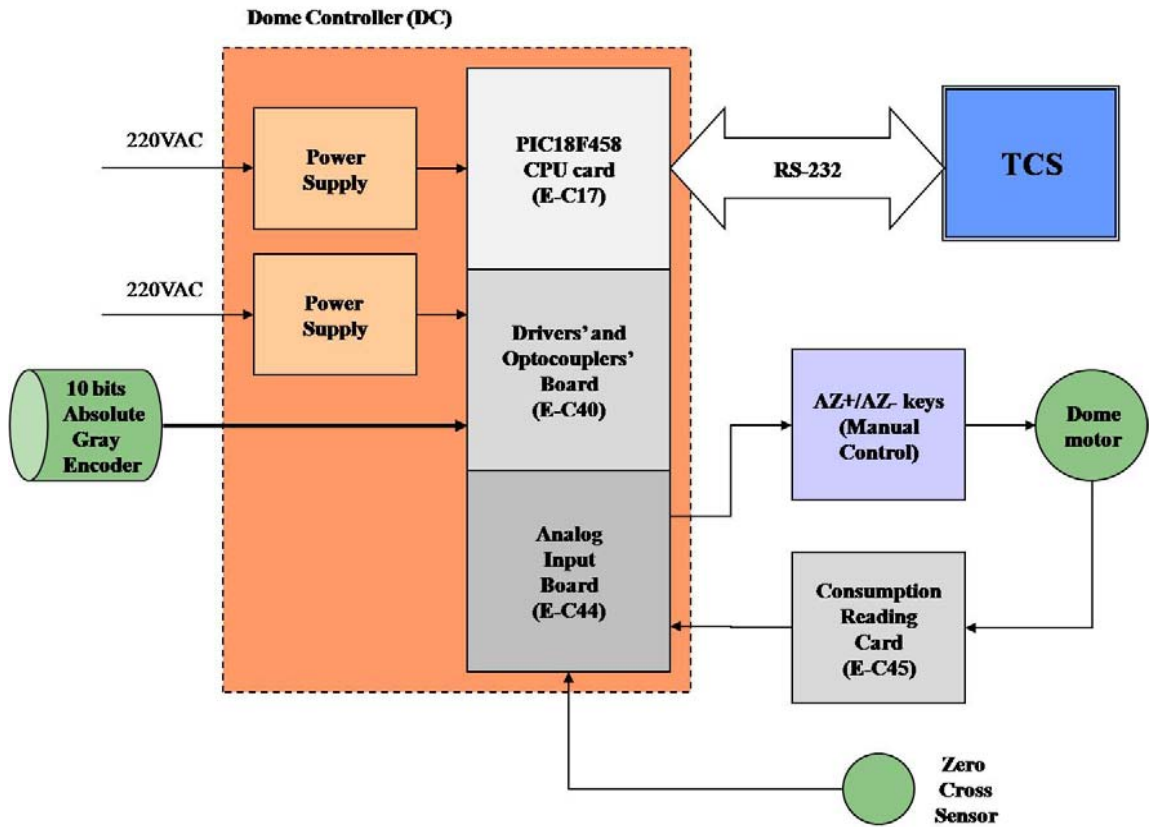

**Figure 4:** Connections among the different elements used in the dome control.

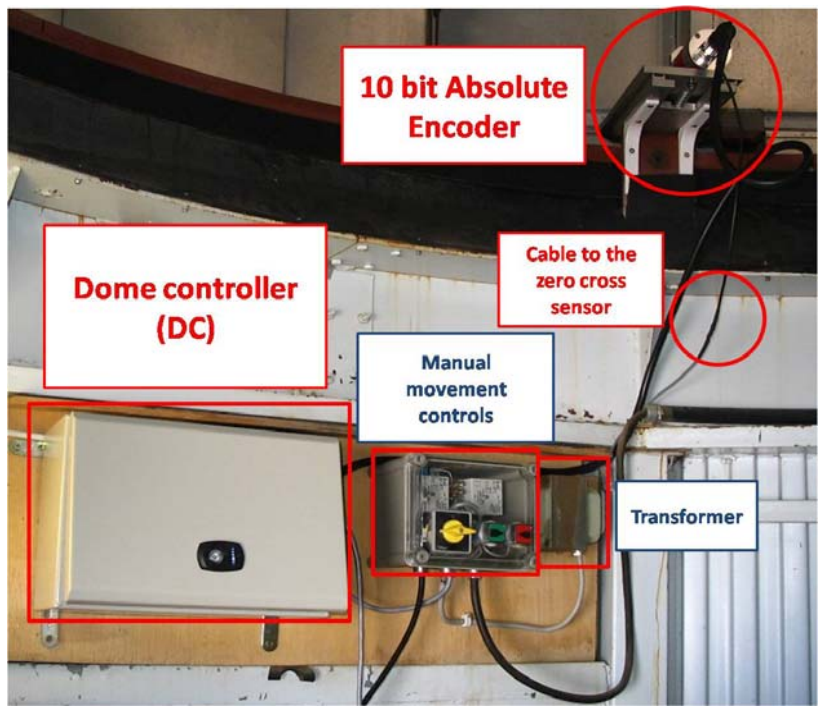

**Figure 5:** Photograph showing part of the system hardware. Red text show elements developed and described in this work.

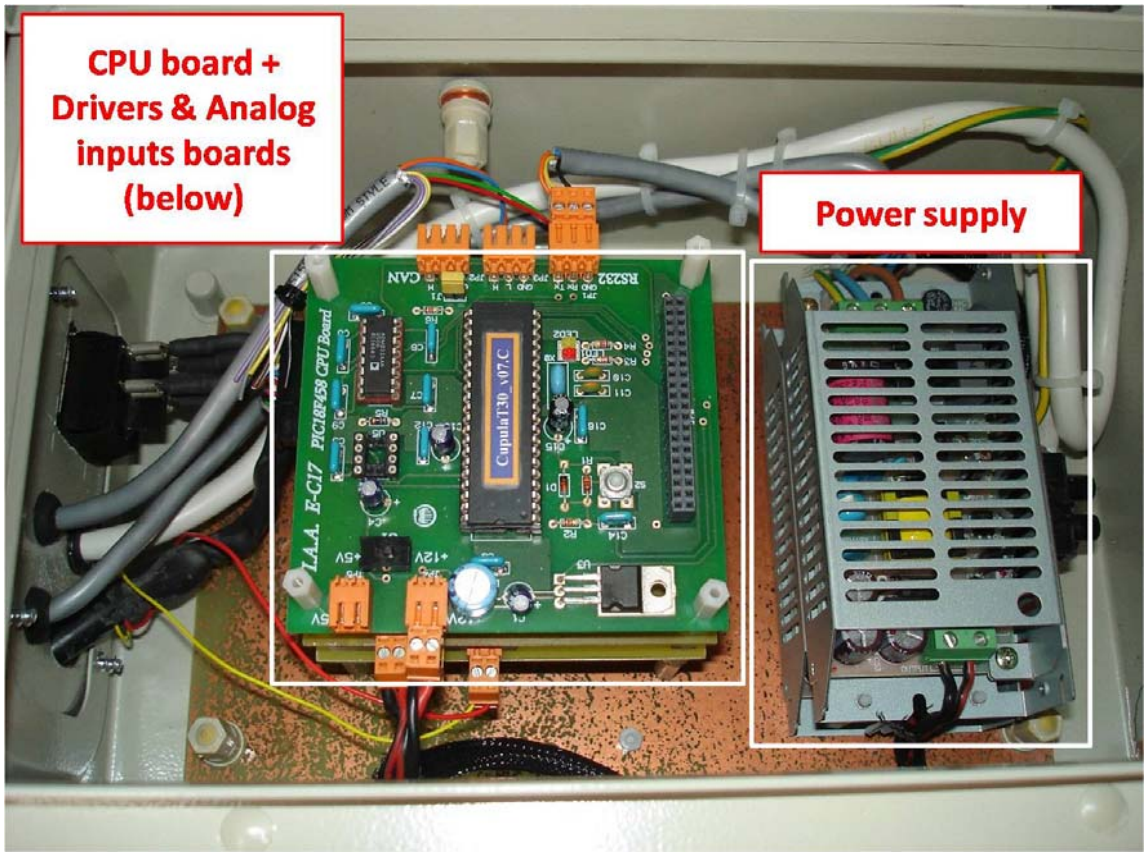

**Figure 6:** Interior of the Dome Controller (DC), showing its different boards and elements.

## 3.2. Software

The Control Software of this automatic telescope is based on the ASCOM (Astronomy Common Object Model) Standards. The telescope, the CCD camera and the filter wheel are ASCOM compliant, so the manufacturers provide the corresponding ASCOM driver interfaces.

Regarding the DC, a program for the 18F458 microcontroller has been developed. The source code has been written in C, using a PICC compiler integrated in the MPLAB environment. The program handles the acquisition of data from the absolute encoder, zero cross sensor and consumption reading card. In response to the read values, it calculates the dome target position and generates the necessary signals for the control of the motor. DC and TCS are connected through an RS-232 link, using control commands and data packets in accordance with a protocol defined by the authors and described in the next section.

The ASCOM Platform includes an application called ASCOM Dome Control Panel, which is a simple Dome control "middleware". It provides a uniform and consistent interface, regardless of the actual hardware and connections used. In our application it was necessary to develop a driver that translates the ASCOM Dome interface to our software RS-232 commands (see section 'ASCOM Dome driver' below).

Although dome control is usually performed through the ASCOM Dome interface, an additional engineering program has been developed in LabVIEW (see Fig. 7). This program will not be running during the usual operation of the system. It only needs to be executed eventually when a closer monitoring of the control and data packets is desired. This result is useful when trying to fix a breakdown, checking the correct operation or implementing new features in the system. As can be seen, the graphic interface displays all the information about the dome status and allows for sending all the control commands. In the upper right corner it displays the present position of the dome, which is updated continuously with the values read from the encoder. In addition to the different buttons that perform the various commands, there is a command line available where a user can write directly the command in the format shown in Table 2. In order to detect interrupts and failures in the communication, all the traffic in the RS-232 link is monitored in the upper right line. Error conditions and zero crosses are also displayed using a pair of LEDs and a text window, where the type of error is displayed.

**Communication protocol**

Communication between TCS and DC is bidirectional, messages structured according the following format:

<STX><command><arguments><CR><LF><ETX>

where <STX>, <CR>, <LF> and <ETX> are the next ASCII control characters:

<STX>: Start of transmission (hexadecimal: 02h)
<ETX>: End of Transmission (03h)
<LF>: Line Feed (0Ah)
<CR>: Carriage Return (0Dh)

In order to make visualization of messages easier, both command and arguments are written in ASCII. <command> has a fixed length of 4 bytes, while <argument> is variable, depending on the command. Table 2 lists all the messages used in the communication, showing the fields <command><argument> but not the control characters. As can be seen, in many of these, <argument> consists of the present or target azimuth, which appear in the table as xxx.

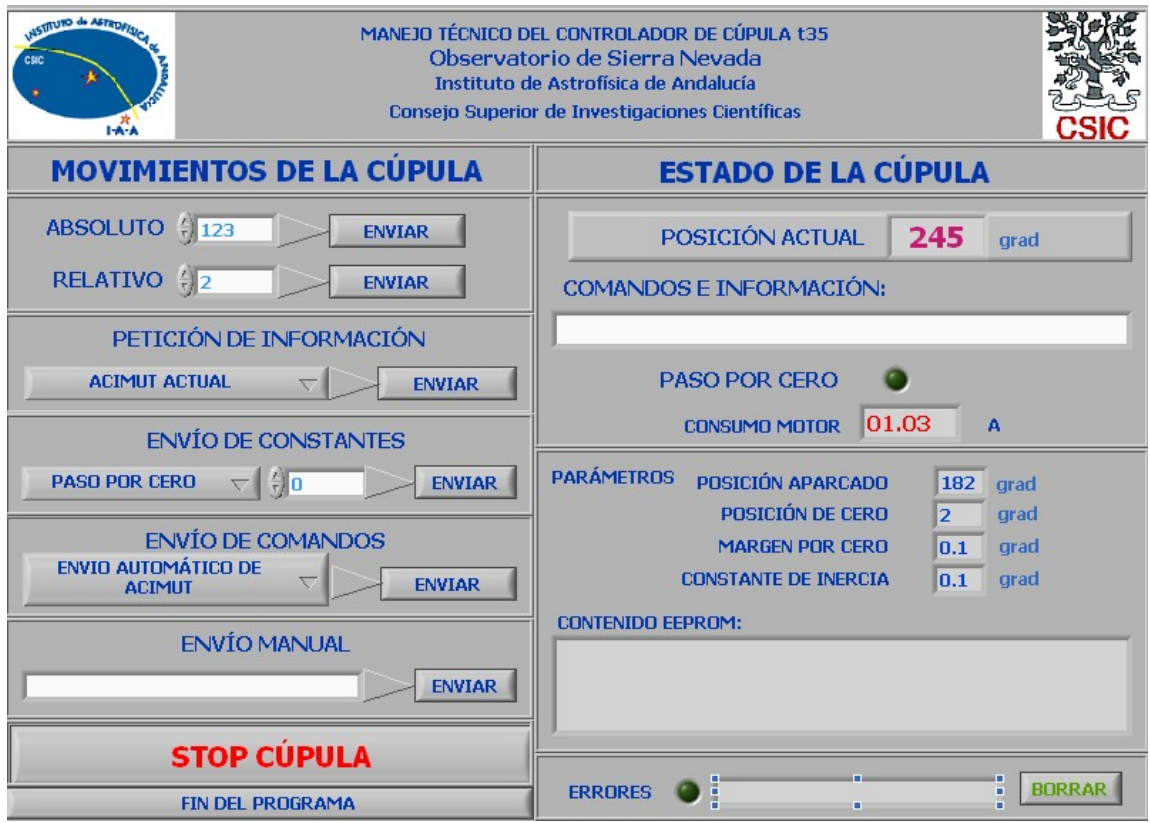

**Figure 7:** Graphic user interface of the engineering program developed for technical purposes.

| Messages from TCS to DC ||
|---|---|
| *Message* | *Meaning* |
| AZABxxx | Absolute movement to azimuth xxx |
| AZR+xxx<br>AZR-xxx | Relative movement (+xxx or –xxx degrees) |
| PARK | Movement to the PARK position |
| STOP | Stops dome movement if it is in motion |
| ¿AZ?<br>¿ST?<br>¿EE?<br>¿CO? | Requires present azimuth, status, EEPROM content, motor consumption. |
| AOFF | Auto messages disabled |
| A_AZ<br>A_CO<br>A_ZE | Auto azimuth, auto consumption, auto zero cross messages enabled. |
| CPARxxx<br>CZERxxx<br>CZEMxxx<br>CINExxx | TCS sends new values for park position, zero cross position, zero margin constant, inertia constant. |

| Messages from DC to TCS | |
|---|---|
| FIN!xxx | Last command executed, dome stopped at azimuth xxx. |
| ¡ZE!xxx | Zero cross detected at azimuth xxx |
| ¡AZ!xxx<br>¡ST!<br>¡EE!<br>¡CO! | Present azimuth, status, EEPROM contents, consumption. These messages are sent in response to ¿AZ? ¿ST? ¿EE? ¿CO? or as automatic information, if the corresponding option is enabled. |
| PAR!xxx<br>ZER!xxx<br>ZERM!xxx<br>INE!xxx | Confirmation of new values for park position, zero position, zero margin or inertia constant, sent by TCS through commands CPARxxx, CZERxxx, CZEMxxx and CINExxx. |
| ERR! | Error message. The type of error is codified in the <argument> field. |

**Table 2:** Description of messages used in the communication between TCS and DC.

**ASCOM Dome driver**

ASCOM is a platform that utilizes a standard interface between astronomic devices and their control software. Any device supplied with an ASCOM driver can be controlled by any ASCOM compliant software. In our application, every element supplied with the telescope (CCD, filter wheel and the telescope itself) included an ASCOM driver. Thus, the main effort in development had to be focused on the dome controller. From the software development point of view, it consisted of programming a dynamic library (DLL) for Windows. Any language suitable for developing Windows objects (COM) can be used. In this application we used C++, in a Microsoft Visual C++ 6.0 compiler.

As we have seen before, communication between the TCS and DC is achieved through a serial link based in the RS-232 standard. Thus, it was also necessary to integrate a serial communication library in the project. We chose for this purpose a freeware code *Serial Communication for WIN32, non-Event driven version* [5] performing the modifications needed to meet our requirements and to follow our serial message format.

Once developed this ASCOM driver, the dome could be controlled through any ASCOM compliant software, like *MaxIm DL*, *TheSky Astronomy Software* or *ACP Observatory Control*. ASCOM platform also provides a small software packet for dome control, called *ASCOM DOME Control Panel*, which is the one we used in our application.

**4. First results**

After installing the new instrument, some spectacular images of the some objects in the sky were taken. Fig. 8 and 9 show images of the IC434 bright red emission nebula around the Horsehead and of the M42 Great Nebula of Orion. The first one is a BVR median combination of 20 x 60 seconds exposures and the second image is a combination of 5 x 60 seconds exposures through BVI filters. Other images can see in the web site http://www.osn.iaa.es/T35/galeria_img.html.

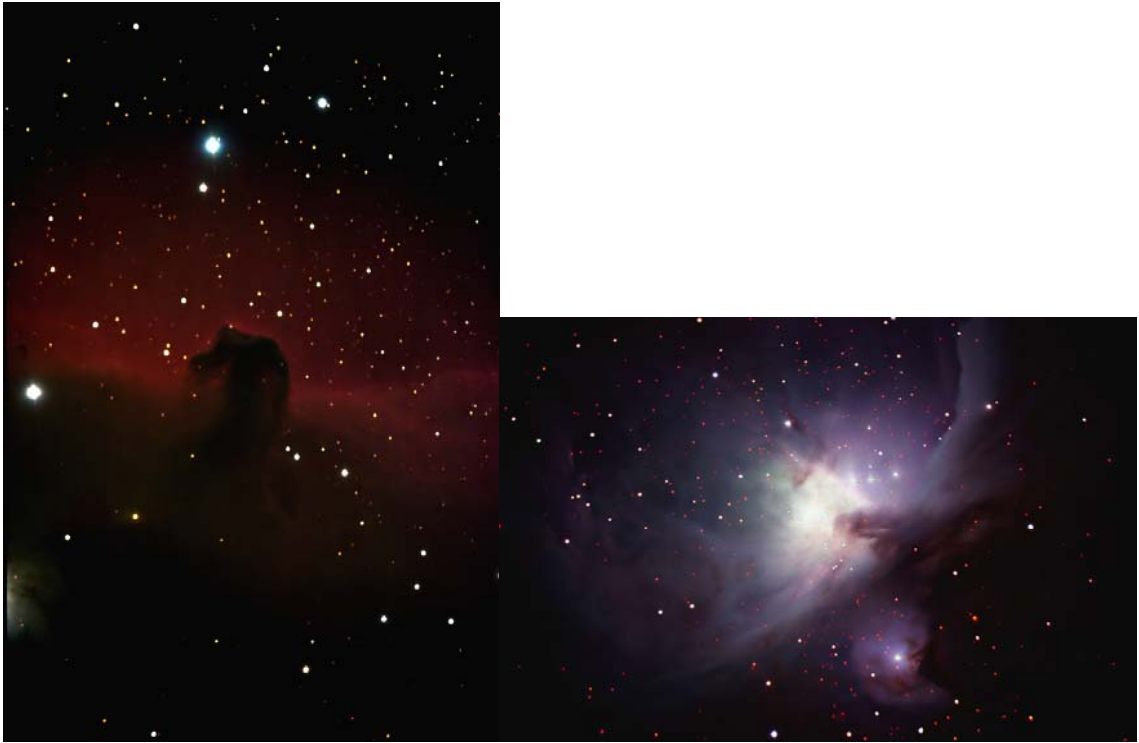

**Figures 8 and 9:** IC 434 and M42 images obtained by using the T35 telescope

Recently, we have carried out several photometric observing campaigns. The IC4756 and NGC7243 open clusters as well as two eclipsing binary systems, HIP7666 and V994Her, were observed during the summer and winter of 2008. This data is being reduced and the results will be published soon.

## 5. Conclusions

We have attained to install and automate a 14'' telescope system in order to perform long-term photometric campaigns. Although the telescope presented some problems of pointing and tracking during the first observations, the behaviour of the telescope and dome control was quite optimum. Both parameters improved fairly when a telescope pointing model was used. The new external system auto-guiding will increase the tracking precision considerably.

The control of the telescope system has been performed through standard ASCOM commands. A simple system based on microprocessor has been designed and implemented allowing the control of the non-standard dome using ASCOM compliant software. Therefore and due to its reliability, this system is being adapted to the two other domes of the SNO.

The control of both telescope and dome can be performed locally or remotely, the latter being the usual mode of operation. The possibility of controlling the system from the facilities of *Instituto de Astrofísica de Andalucía*, in the town of Granada, implies an important logistic advantage and considerable time saving.

**Ackowledgment**

We would like to acknowledge the staff in SNO for his technical, logistic and human support. SMR thanks Victor Costa, Rafael Garrido, José Luis Ortiz and Lourdes Verdes- Montenegro for their useful support.